\documentclass[prc,twocolumn,floatfix,amsmath,superscriptaddress]{revtex4-2}

\usepackage{graphicx}% Include figure files
\usepackage{dcolumn}% Align table columns on decimal point
\usepackage{bm}% bold math
\usepackage{xcolor}
\usepackage{url}       % for basic URL support
\usepackage{hyperref}  % for clickable URLs
\usepackage{amsmath,amssymb}
\usepackage{booktabs, multirow}
\usepackage{siunitx}

\begin{document}
\title{
The recoil nucleon effect in nuclear systems}

\author{Han Kui Wang}
\email{whk@zstu.edu.cn  }
\affiliation{Department of physics, Zhejiang Sci-Tech
University, Hangzhou 310018, China}

%\author{Baohua Sun}
%\affiliation{School of Physics, Beihang University, 100191, beijing, China}

%\author{Yang Sun}
%\affiliation{School of Physics and Astronomy, Shanghai Jiaotong University, 200240, Shanghai, China}

\date{\today}% It is always \today, today,
             %  but any date may be explicitly specified

\begin{abstract}

Nuclear excitation energies are typically in the keV-MeV range,  but rare eV-scale transitions remain a puzzle for nuclear theory.
This work develops a two-body recoil  model to describe $\beta^-$ decay on the basis of classic momentum conservation.
Using the final states of $\beta^-$ decay with a keV-scale excitation energy, 
the model yields eV-scale recoil energy for a single nucleon.
It indicates the recoil nucleon effect in nuclear systems:
a single nucleon bound in a nucleus can be excited by absorbing the recoil momentum.
 % emitted in the $\gamma$ decay.
The observed eV-scale isomers in $^{235}$U and $^{229}$Th can be attributed to the recoil neucleon effect.
This effect could spark a eV-scale energy, and
provide a solid theoretical foundation for predicting such extremely small energies.
The new discovery will place low-energy nuclear isomer research at the cutting edge of new-physics searches, 
while offers a clean and complementary alternative to high-energy accelerator experiments.

\end{abstract}

%\keywords{Suggested keywords}%Use showkeys class option if keyword
                              %display desired
\maketitle

\section{Introduction}

Energy, the fundamental driver of physical phenomena, manifests across a distinct hierarchy of scales, each dominated 
by characteristic processes.
Chemical bonds and molecular transformations involve energies on the order of a few electronvolts  (eV)~\cite{ref1}.
Atomic ionization energies typically range from a few eV to several tens of eV ~\cite{ref2}.
Nuclear excitation energies are conventionally located at kiloelectronvolt (keV) to megaelectronvolt (MeV) scales ~\cite{ref3}, while nuclear reactions and fundamental particle interactions release energies from MeV up to gigaelectronvolt (GeV) levels and beyond~\cite{ref4,ref5}.
This well-established energy hierarchy governs the structure of matter and cosmic evolution, from molecular dynamics to stellar nucleosynthesis and high-energy astrophysical phenomena.

The existence of  eV-scale energy level in nucleus defies the typical nuclear level systematics ~\cite{Th229-1, U235-1}.
Early estimates placed the transition wavelength near 149.7(31) nm, 
making it accessible via vacuum-ultraviolet (VUV) laser sources based on high-harmonic generation ~\cite{Nature573}.
A major milestone was achieved with the first direct observation of the radiative decay of $^{229m}$Th in a solid-state host, 
where VUV spectroscopy of thorium implanted in CaF$_2$ and MgF$_2$ crystals at the CERN-ISOLDE facility yielded a transition energy of 8.338(24) eV and a half-life of 670(102) s in MgF$_2$~\cite{Nature617}.
Ref. ~\cite{nature2024} establishes a direct frequency link between the $^{229}$Th isomeric transition and an atomic clock, By referencing a VUV frequency comb to the JILA Sr87 optical clock, yielding a nuclear transition frequency of 2,020,407,384,335(2) kHz.

The modern pursuit of a nuclear clock based on the eV-scale transition is pioneered by Peik and Tamm, 
who theoretically investigated laser excitation of the $^{229}$Th nucleus and outlined its potential for a clock with unprecedented accuracy and sensitivity to variations of the fine-structure constant~\cite{EPL61}.
Such a clock, operating via nuclear transitions rather than electron-shell dynamics ~\cite{PRL108}, would facilitate unprecedented tests of gravitational redshift in relativistic geodesy~\cite{RPP81},
enhanced sensitivity to dark-matter interactions ~\cite{Nat-Phys10,PRX15},
and Searches for temporal variations in fundamental constants ~\cite{PRL97}.

Nuclear excitation energies are typically higher than a few keV.
Yet, a few remarkable exceptions at the eV scale fall drastically outside this scheme, 
posing a profound challenge to our understanding of nuclear structure.
They are the ~8.36 eV level in $^{229}$Th,
the ~76.7 eV level in $^{235}$U,
and the possible one suggested below 300 eV in $^{73}$Ga~\cite{Ga73-3}.
The origin of these eV-scale levels cannot be explained by nuclear models alone,
nor by atomic contributions such as the $M\ddot{o}ssbauer$  effect~\cite{mse-1958}.
%While nuclear models struggle to reproduce such low excitation energies, electron effects are
%orders of magnitude too weak to account for them.
This work will develop
a Two-Body Recoil (TBR) model using momentum conservation  to explain such extremely small excitation energies in nuclear sysytems.

\section{The Two-body Recoil Model}

\iffalse
Consider an object (or particle)  of mass $M + \delta m$ at rest, with total momentum zero in the initial STATE.
In the final STATE, it breaks into a larger fragment of mass M and a smaller one $\delta m$.
The larger fragment has momentum $MV$ and recoil kinetic energy $E_R$, 
while the mass  $\delta m$ has momentum $\delta m v$ and kinetic energy $E_x = 1/2 \delta m v^2$.
Using conservation of momentum,  $MV = \delta m v$,
one obtains
the recoil kinetic energy $E_R$:
\begin{eqnarray}
E_R = \frac{1}{2} MV^2  = \frac{1}{2} M (\frac{\delta m v}{M})^2
%=  \frac{\delta m }{M} (\frac{1}{2} \delta m v^2)
=  \frac{\delta m }{M} E_x
\end{eqnarray}

Let us use the recoil momentum model to simulate the Beta decay process.
Beta decay is a type of radioactive decay where the unstable atomic nucleus transforms 
by capturing an electron or emitting a beta particle (an electron or positron) to achieve greater stability.
For Beta minus decay,
a neutron in the nucleus converts into a proton, while an electron and an electron antineutrino are emitted.
\begin{eqnarray}
n  \rightarrow p + e^- + \nu^-_e  \nonumber
\end{eqnarray}
If only mind the initial and final state of beta minus decay process,
one can set the intial intial STATE of the recoil momentum model as a neutron (mass $m_n = M + \delta m$ )  with zero meomentum,
and final STATE as a proton (mass $m_p$ = M ), 
with the momentum treated as equivalent to that obtained from the kinetic energy $E_x$ of the  mass $\delta m$.
\fi

$\beta$ decay is a radioactive process in which an unstable nucleus emits a $\beta$ particle (an electron $e^-$ or a positron $e^+$) 
together with a corresponding neutrino or antineutrino, thereby moving toward a more stable proton-to-neutron ratio. 
In the case of $\beta^-$ minus decay, 
a neutron inside the nucleus converts into a proton, emitting an electron and an electron antineutrino:
\begin{eqnarray}
n  \rightarrow p + e^- +\overline{\nu}_e  \nonumber
\end{eqnarray}

Let us develop a two-body recoil model to describe $\beta$ decay.
Consider a parent particle of mass $M + \delta m$ initially at rest, so that its total momentum is zero. 
In the final STATE, it breaks into two fragments: a heavier one of mass M and a lighter one of mass $\delta m$.
The heavier fragment recoils with momentum $MV$ and recoil kinetic energy $E_R$, 
while the lighter fragment $\delta m$ carries momentum $\delta mv $ and kinetic energy $E_x = 1/2 \delta m v^2$.
By conservation of momentum, we have $MV = \delta m v$. 
Consequently, the recoil energy  $E_R$ can be expressed as
\begin{eqnarray}
E_R = \frac{1}{2} MV^2  = \frac{1}{2} M (\frac{\delta m v}{M})^2
%=  \frac{\delta m }{M} (\frac{1}{2} \delta m v^2)
=  \frac{\delta m }{M} E_x
\end{eqnarray}

Focusing on a single nucleon in the $\beta^-$ decay process from $^A_Z X_N$ to $^A_{Z+1} X_{N-1}$, while keeping the core  $^{A-1}_{Z} X_{N-1}$ frozen,
allows this process to be mapped onto the two-body recoil model described above.
Here, X denotes the element symbol, A the nucleon (mass) number, Z the proton number, 
and N the neutron number.
The initial STATE of the TBR model can be taken to be a neutron at rest,with mass $m_n = M + \delta m$.
The final STATE is a proton of mass \( m_p = M \),
with the light decay products (electron and antineutrino) collectively 
treated as a light fragment of mass $\delta m$,
which serves as the equivalent momentum carrier.
The momentum of the mass $\delta m$ is equal in magnitude to the recoil momentum of the proton.
The equivalent kinetic energy  $E_x(\delta m)$ can be directly taken to be the excitation energy of the final state in the  $\beta^-$ decay.

In the final STATE of the model,
the  mass $\delta m$ is defined in two ways:
$\delta m'$, excluding $m_e$, i.e., for a static electron with no kinetic energy, is
\begin{eqnarray}
\delta m = m_n - m_p - m_e = 0.782 MeV/c^2;
\end{eqnarray}
$\delta m$, including the electron mass $m_e$ (WI), is
\begin{eqnarray}
\delta m' = m_n - m_p = 1.293 MeV/c^2
\end{eqnarray}
Here, $\delta m$ denotes the neutron decay energy, and $\delta m'$ corresponds to the neutron-proton mass difference.
For a fixed $E_x$, with $ \delta m = 0.782 MeV/c^2$ , one obtains
a lower bound on the calculated recoil energy:
\begin{eqnarray}
E_R = \frac{0.782 MeV}{m_p} E_x
\end{eqnarray}
With $ \delta m = 1.293 MeV/c^2$ , one obtains
the upper bound of the recoil energy:
\begin{eqnarray}
E'_R = \frac{1.293 MeV}{m_p} E_x
\end{eqnarray}

\section{The recoil nucleon effect}

With the two-body recoil model established, 
we now apply it to a specific $\beta^-$ decay from $^{235}Pa$ to $^{235}U$, focusing on a single nucleon while keeping the $^{234}Pa$ core frozen.
The initial STATE is taken to be a neutron at rest in the ground state of $^{235}Pa$.
The final STATE contains a proton in $^{235}U$, 
together with a light fragment of mass $\delta m$ that 
serves as an equivalent momentum carrier for the emitted leptons.
The recoil momentum of the proton is equal in magnitude to the total momentum carried away by the light fragment $\delta m$. 
Using the equivalent kinetic energy $E_x$ = 13.1 keV in Eq.(4) yields 
the recoil energy:
 \[
E_R = 10.9~eV. 
\]
The 13.1-keV energy is from the 3/2$^+$ level in $^{235}U$, 
which is taken to be the final state of the
 $\beta^-$  decay from the $(3/2^-)$ initial state in $^{235}Pa$.
Using the $E_x$ = 129.3 keV associated with another final state of the $\beta^-$ decay yields 
 \[
E_R = 107.8~eV. 
\]

For the $\beta^-$ decay of $^{229}Ac$ to $^{229}Th$, keeping the  $^{228}Ac$ core frozen,
the initial STATE is taken to be a neutron at rest in the $^{229}Ac$ ground state.
The final STATE in $^{229}Th$ comprises a proton plus a mass $\delta m$ acting as 
an equivalent momentum carrier.
The equivalent kinetic energy $E_x$ can be taken to be the excitation energy of the final state of the $\beta^-$ decay,
for examples, the $5/2^+$ level  at 29.19 keV and the  $5/2^-$ one at 146.35 keV.
Using $E_x$ = 29.19 keV in Eq.(4) yields 
the recoil energy:
 \[
E_R = 24.3~eV. 
\]
Using $E_x$ = 146.35 keV in Eq.(4) yields 
the recoil energy:
 \[
E_R = 121.5~eV. 
\]

It appears that, using final state of $\beta^-$ decay  at keV-scale energy, 
the TBR model yields eV-scale recoil energy carried by a single nucleon. 
If a single nucleon inside nucleus is able to absorb the recoil momentum,
we shall calculate its recoil energy in a $\gamma$ decay.
The recoil momentum $P_R$ associated with a $\gamma$-ray transition is:
\begin{equation}
P_R = E_\gamma/c       \nonumber
\end{equation}
Here $E_\gamma$ is the energy of a photon releasing in $\gamma$ decay, 
while contant c is the light speed.
In the non-relativistic limit ($v_R \ll  c $), its recoil kinetic energy is 
\begin{equation}
E_R = \frac{P_R^2}{2m_N} = \frac{E_\gamma ^2}{2m_N c^2}.
\end{equation}
Here, $m_N$ is the mass of a nucleon.

\iffalse
Solving for the photon energy, we obtain:
\begin{equation}
E_\gamma = \sqrt{2E_R m_N c^2}
\end{equation}
\fi
\iffalse
The  energy  381 keV owes to the $\gamma$ decay from the 7/2+ level  at 393.9 keV to 
the 3/2+ level  at 13.1 keV.

The 381 keV  $\gamma$ ray arises from the decay of the 7/2+ level  at 393.9 keV to 
the 3/2+ level  at 13.1 keV.

It is a mazing discovery that a single nucleus inside atom can be excited by absorbing the recoil momentum emitted by 
gamma decay, so called the $M\ddot{o}ssbauer$  effect.

\fi

Using $E_\gamma$ = 381 keV in formula (6) yields 
\[
E_R =  \frac{E_\gamma ^2}{2m_n c^2} = 77.4~eV.
\]
Here, $m_N$ is selected as the neutron mass $m_n = 939.565~MeV/c^2$.
This 381 keV transition is attributed to the $\gamma$ ray from the 7/2$^+$ level  at 393.9 keV to 
the 3/2$^+$ level at 13.1 keV.
Notably, the calculated recoil energy of 77.4 eV is nearly equal to
the isomer energy of 76.737(18) eV in $^{235}U$ ~\cite{U235-2}.

In $^{229}Th$, 
taking the excited energy $E_\gamma$ = 125.4 keV of the 7/2$^+$ level  in Eq.(6)
yields the recoil energy:
\[
E_R =  \frac{E_\gamma ^2}{2m_n c^2} = 8.38~eV.
\]
This calculated value is almost identical to the eV-scale isomer in $^{229}$Th, namely 8.355733~eV measured in Ref.~\cite{8.355733}.
The two precise correspondences indicate that the observed eV-scale isomers 
can be attributable to the recoil energy $E_R$ of a single nucleon.
It is an amazing discovery that a single nucleon bound in a nucleus can be excited by absorbing the recoil momentum emitted in 
the gamma decay, named the reCoil Nucleon (CN) effect.

By comparison with the $M\ddot{o}ssbauer$  effect~\cite{mse-1958},
when a nucleus is bound in a crystal lattice,
the recoil momnetum associated with $\gamma$-ray emission is effectively transferred to
the entire crystal.
Since the crystal mass is essentially infinite, the recoil energy is virtually zero.
If the recoil momentum from the $\gamma$ transition of energy $E_\gamma = 125.4 $ keV were absorbed 
by a nucleus rather than by the crystal,
the recoil energy  would be:
\[
E^{\gamma}_R = \frac{E_\gamma ^2}{2m(U^{235}) c^2} = 0.036~eV  \nonumber
\]
where the effective recoil mass is the whole  $^{235}$U nucleus, with $m(U^{235}) c^2 \approx  218940 MeV/c^2$.

For check the CN effect, let us see 
the third possible eV-scale excited energy in nucleus $^{73}$Ga.
Ref.~\cite{Ga73-3} reports a energy separation  is less than 300 eV between the 3/2- level and 1/2- ground state.
Using  $E_\gamma$ = 495.7 keV in formula (6),
yields the recoil energy: 
$E_R$ = 130.9 eV.
The energy 495.7 keV is from the final state of $\beta$ decay at 495.93 keV in $^{73}$Ga, 
which wil decay to the very 3/2- level  mentioned above.
In fact, taking $E_\gamma < $  750 keV can result in a recoil energy below 300 eV.

Above all, 
the eV-scale excited energy inside nucleus should be attributabled to the recoil nucleon effect.
A single nucleon bound in nucleus is able to be exicted, for example, absorbing the recoil momentum from $\gamma$ decay, and sparks very tiny excited energy in nuclear world.

\iffalse
A:
The energy 495.7 keV owes to the final state at 495.93 keV in $\beta$ decay, which decays to the very level 3/2- mentioned above.
In fact, taking $E_\gamma$ less than 750 keV could obtain a recoil energy less than 300 eV.
B:
The energy 495.7 keV is attributed to the final state at 495.93 keV in $\beta$ decay, 
which then decays to the very 3/2- level  mentioned above.
In fact, taking $E_\gamma < $  750 keV can result in a recoil energy below 300 eV.

Applying the Eq.~(7) with $E_R$ = 76.7 eV and $m_n = 939.565 MeV/c^2$ yields 
a calculated  gamma decay energy of 
\[
E^{th}_\gamma = 378  keV.
\]
This value is close to the observed transition energy of 367 keV,
which corresponds to an E1 decay from the  $J^\pi = 7/2^+$ level directly to the ground state. 
Using the experimental value $E_\gamma = 367 $ keV in Eqs.(6),
the recoil energy imparted to a single neutron is obtained as 
\[
E_R = \frac{E_\gamma ^2}{2m_n c^2} = 71.37 eV,  
\]
which agrees very well with the 76.7 eV  isomer in 235U.

\fi

\iffalse
before:
This value is closed to the observed decay energy 367 keV from  the level  $J^\pi = 7/2^+$ 
directly to the ground state with $E_1$ transition.

Using the experimental value $E_\gamma = 367 $ keV in Eqs.(6),
one gets the recoil energy carried by a single neutron:

This value is very close to the isomer state 76.7 eV of 235U.
modified:

It seems
the CN effect explains very well the eV-scale excited energy 76.7 eV observed in $^{235}$U.
This isomer should be excited by the single nucleon absorbing the recoil momentum from beta or gamma decay.
\fi

\section{Conclusions}

\iffalse
According to Mossbauer effect, atomic nuclei embedded in a crystal or target behave as recoil nuclei
with infinite mass, and thus experience no recoil effect.
The recoil nucleon effect shows that the weak interaction can 'tear open' the nucleus dominated by the strong interaction, liberate single-nucleon degrees of freedom, and thereby enable beta decay.

The explanation for the 8.36 eV level in \(^{229}\text{Th}\) or the 76.7 eV  level in \(^{235}\text{U}\) relies heavily on fortuitous single-particle orbital proximity and delicate shape deformations.

The recoil nucleon effect shows that the weak interaction can 'tear open' the nucleus dominated by the strong interaction, liberate single-nucleon degrees of freedom, and thereby enable beta decay.
\fi

A two-body recoil model is developed to describe the $\beta^-$ decay using momentum conservation.
The model predicts eV-scale recoil energy for the proton 
when the equivalent kinetic energy is taken to be a keV-scale energy
 of the final state in the $\beta^-$ decay.  
This model indicates the recoil nucleon effect in nuclear world:  
a single nucleon can be excited by absorbing the recoil momentum from $\beta$ decay or $\gamma$ decay.
This effect unifies the origin of the eV-scale excited energies in $^{235}$U, $^{229}$Th and $^{73}$Ga
as a simple and compelling physical mechanism.
The singular \(^{229}\text{Th}\) isomer presents experimental challenges such as vacuum-ultraviolet accessibility and environmental sensitivity.
The recoil neucleon effect will provide a solid theoretical basis for predicting such eV-scale levels, 
and push low-energy nuclear isomer experiments to the forefront of new-physics searches,
while offering a clean and complementary probe to high-energy accelerator experiments.

\section{acknowledgements}

Research at ZSTU was supported by the National Natural Science Foundation of China (Grants No.12475124, No.U2267205), and ZSTU intramural grant(22062267-Y).

\bibliography{NM-1}% Produces the bibliography via BibTeX.

\begin{figure*}
\centering
\includegraphics[width = 12 cm]{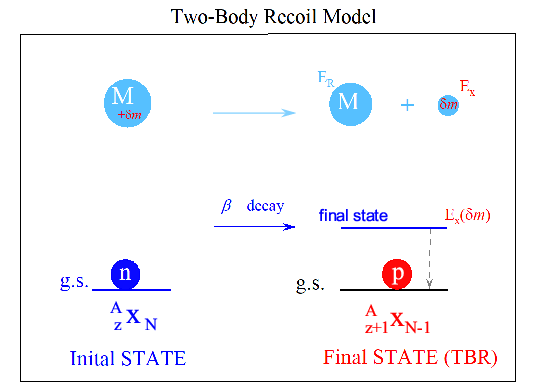}
\caption{The two-body recoil model.
\iffalse
$E_R$ notes the recoil energy of mass M.
$E_x$ notes the kinetic energy of mass $\delta m$.
"n" labels neutron, while "p" labels proton.
$^A_Z X_N$ notes nucleus with mass A, proton number Z and neutron number N.
\fi
Let $E_R$ denote the recoil energy of mass M, and 
$E_x$  the kinetic energy of mass $\delta m$.
The labels n and p denote a neutron and a proton, respectively.
Let $^A_Z X_N$ denote nucleus with mass A, proton number Z and neutron number N.
}
\end{figure*}

\begin{figure*}
\centering
\includegraphics[width = 14 cm]{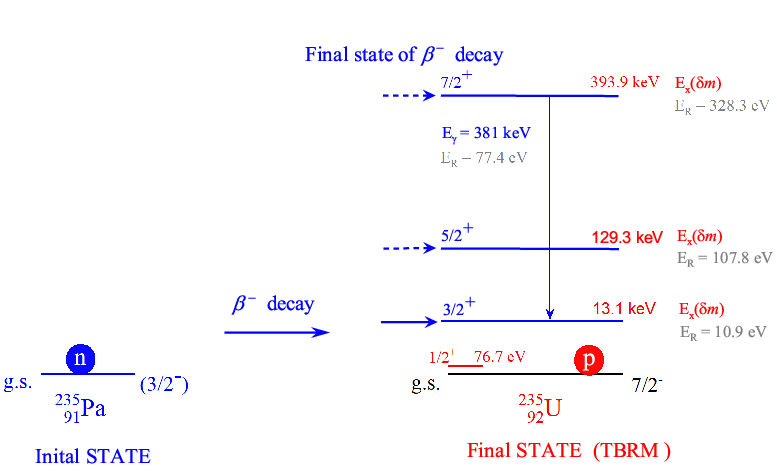}
\caption{ 
The two-body recoil model applied to the $\beta^-$ decay of $^{235}$Pa to $^{235}$U.
The equivalent energy $E_x(\delta m)$ takes the excitation energy of the final state in $\beta^-$ decay, 
and $E_R$ denotes recoil energy.
}
\end{figure*}

\begin{figure*}
\centering
\includegraphics[width = 14 cm]{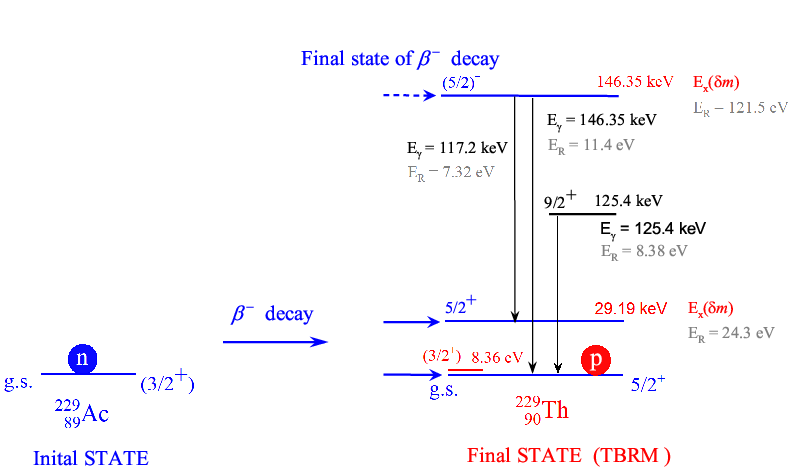}
\caption{
Two-body recoil model applied to the $\beta^-$ decay of $^{229}$Ac to $^{235}$Th.
The equivalent energy $E_x(\delta m)$ takes the excitation energy of the final state in $\beta^-$ decay, 
and $E_R$ denotes recoil energy.
}
\label{Th229}
\end{figure*}

\end{document}